# The Fast Haar Wavelet Transform for Signal & Image Processing

V.Ashok
Department of BME,
Velalar College of Engg.&Tech.
Erode, India – 638012.

T.Balakumaran
Department of ECE
Velalar College of Engg.&Tech
Erode, India – 638012

C.Gowrishankar
Department of EEE
Velalar College of Engg.&Tech
Erode, India – 638012

Dr.ILA.Vennila
Department of ECE,
PSG College of Technology,
Coimbatore,
TamilNadu, India

Dr.A.Nirmal kumar
Department of EEE,
Bannari Amman Institute of Technology,
Sathyamangalam,
TamilNadu, India

*Abstract*- **A method for the design of Fast Haar wavelet for signal processing & image processing has been proposed. In the proposed work, the analysis bank and synthesis bank of Haar wavelet is modified by using polyphase structure. Finally, the Fast Haar wavelet was designed and it satisfies alias free and perfect reconstruction condition. Computational time and computational complexity is reduced in Fast Haar wavelet transform.**

## I. INTRODUCTION

The wavelet transform has emerged as a cutting edge technology, within the field of signal & image analysis. Wavelets are a mathematical tool for hierarchically decomposing functions. Though routed in approximation theory, signal processing, and physics, wavelets have also recently been applied to many problems in computer graphics including image editing and compression, automatic level-of-detailed controlled for editing and rendering curves and surfaces, surface reconstruction from contours and fast methods for solving simulation problems in 3D modeling, global illumination and animation [1].

Wavelet theory was developed as a consequence in the field of study the multi-resolution analysis. Wavelet theory can determine the nature and relationship of the frequency and time by analysis at various scales with good resolutions.

Time-Frequency approaches were obtained with the help of Short Time Fourier Transform (STFT). For the better time (or) frequency resolution (but not both) can be determined by individual preference (or) convenience rather than by necessity of the intrinsic nature of the signal, the wavelet analysis gives the better resolution [2].

According to the applications, the biomedical researchers have large number of wavelet functions from which to select the one that most closely fits to the specific application. Wavelet theory has been successfully applied to a number of biomedical problems [3-5]. Many applications such as image compression, signal & image analysis are dependent on power availability. In this paper, a method for design of Haar wavelet for low power application is proposed. The main idea of this proposed method is the decimated wavelet coefficients are not computed. This makes the conservation of power and reduces the computation complexity. The Haar wavelet which makes the low power design is simple and fast. The proposed design approach introduces more savings of power.

This paper organised as follows. In Section II, the existing Haar wavelet is introduced. In section III presents Haar wavelet analysis bank reduction. In section IV presents Haar wavelet synthesis bank reduction. In section V presents Haar wavelet and Fast Haar wavelet experimental results are shown as graphical output representation to the signal and image processing and we conclude this paper with section VI.

## II. HAAR WAVELET STRUCTURE

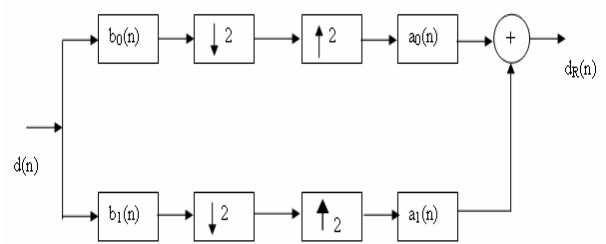

Fig. 1 Two channel wavelet structure

The wavelet transform can be implemented by a two channel perfect reconstruction (PR) filter bank [6]. A filter bank is a set of filters, which are connected by sampling operators. Fig.1 shows an example of a two-channel filter bank applied by one dimensional signal. $d(n)$ is an input signal and $d_R(n)$ is reconstructed signal. In the analysis bank, $b_0(n)$ is a analysis low pass filter and $b_1(n)$ is a analysis high pass filter. However in practice, the responses overlap, and decimation of the sub-band signals, which are results in aliasing. The fundamental theory of the QMF bank states that the aliasing in the output signal $d_R(n)$ can be completely canceled by the proper choice of the synthesis bank [7]. In the synthesis bank, $a_0(n)$ is the reconstruction low pass filter(LPF) and $a_1(n)$ is the reconstruction high pass filter (HPF). Low pass analysis coefficients of Haar Wavelet is $\left[\frac{1}{\sqrt{2}} \ \frac{1}{\sqrt{2}}\right]$. High pass analysis coefficients of Haar Wavelet is $\left[-\frac{1}{\sqrt{2}} \ \frac{1}{\sqrt{2}}\right]$. Low pass synthesis coefficients of Haar Wavelet is $\left[\frac{1}{\sqrt{2}} \ \frac{1}{\sqrt{2}}\right]$. High pass synthesis coefficients of Haar Wavelet is $\left[\frac{1}{\sqrt{2}} \ -\frac{1}{\sqrt{2}}\right]$.

### III. HAAR WAVELET ANALYSIS BANK REDUCTION

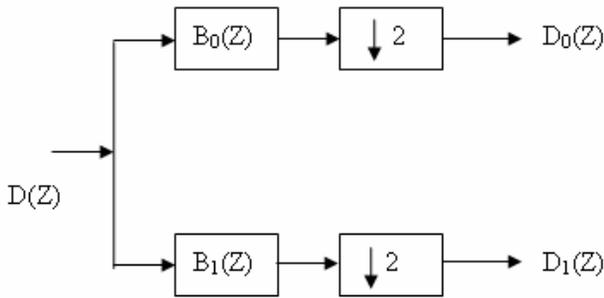

Fig. 2 Analysis bank of wavelet structure

Fig.2 shows analysis bank of wavelet structure. $d(n)$ is an input signal, $d_0(n)$ is an low pass output of $d(n)$ and $d_1(n)$ is high pass output of input signal.

For simplicity write in Z domain

$$D_0(Z) = \tfrac{1}{2}[D(Z^{1/2})B_0(Z^{1/2}) + D(-Z^{1/2})B_0(-Z^{1/2})] \quad (1)$$

$$D_1(Z) = \tfrac{1}{2}[D(Z^{1/2})B_1(Z^{1/2}) + D(-Z^{1/2})B_1(-Z^{1/2})] \quad (2)$$

At Perfect Reconstruction condition, No Aliasing Components presents

$$D_0(Z) = \tfrac{1}{2}[D(Z^{1/2})B_0(Z^{1/2})] \quad (3)$$

$$D_1(Z) = \tfrac{1}{2}[D(Z^{1/2})B_1(Z^{1/2})] \quad (4)$$

From Quadrature Mirror Filter by [7], analysis filters are chosen as follows

$$B_0(Z) = B(Z) \leftrightarrow b(n) \quad (5)$$

$$B_1(Z) = B(-Z) \leftrightarrow (-1)^n b(n) \quad (6)$$

Transfer function $B(Z)$ of an LTI system can decomposed into its polyphase components [9].

$B(Z)$ can be decomposed into

$$B_0(Z) = \sum_{\lambda=0}^{M-1} Z^{-\lambda} B_\lambda(Z^M) \quad (7)$$

In Haar Wavelet M=2

So Low pass filter & High pass filter is

$$B_0(Z) = B_{00}(Z^2) + z^{-1}B_{01}(Z^2) \quad (8)$$

$$B_1(Z) = B_{00}(Z^2) - z^{-1}B_{01}(Z^2) \quad (9)$$

Sub $B_0(Z)$, $B_1(Z)$ in Eq (3) & (4)

$$D_0(Z) = \tfrac{1}{2}[D(Z^{1/2})(B_{00}(Z) + z^{-1/2}B_{01}(Z))]$$

$$D_0(Z) = \tfrac{1}{2}[D(Z^{1/2})(B_{00}(Z) + \tfrac{1}{2}z^{-1/2}D(Z^{1/2})B_{01}(Z)] \quad (10)$$

In Haar wavelet $B_{00}(Z) = B_{01}(Z)$

$$D_0(Z) = B_{00}(Z)[\tfrac{1}{2}D(Z^{1/2}) + \tfrac{1}{2}Z^{-1/2}D(Z^{1/2})] \quad (11)$$

Like

$$D_1(Z) = D(Z^{1/2})B_{00}(Z) - \tfrac{1}{2}Z^{-1/2}D(Z^{1/2})B_{01}(Z)$$

$$D_1(Z) = B_{00}(Z)[\tfrac{1}{2}D(Z^{1/2}) - \tfrac{1}{2}Z^{-1/2}D(Z^{1/2})] \quad (12)$$

Combining Eq (11) & (12)

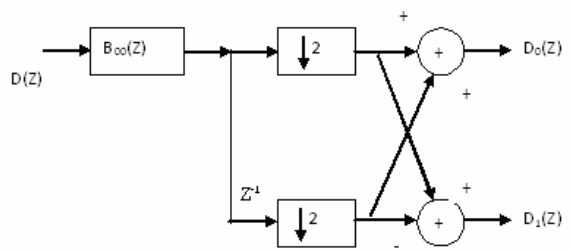

Fig. 3 Modified analysis bank structure

$B_{00}(Z) \longleftrightarrow b_{00}(n)$. In haar wavelet $b_{00}(n) = \frac{1}{\sqrt{2}}$.

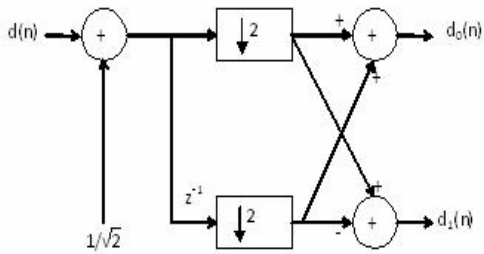

Fig. 4 Fast Haar wavelet analysis bank

Shifting the down sampler to the input bring reduction in the computational complexity of factor 2 along with it. Fig.4 shows Fast Haar wavelet analysis structure compared to original Haar wavelet structure, Number of arithmetic calculations are reduced in Fast Haar wavelet structure. But using above method computational complexity [10] reduced in less than quarter of original computational complexity.

## IV. HAAR WAVELET SYNTHESIS BANK REDUCTION

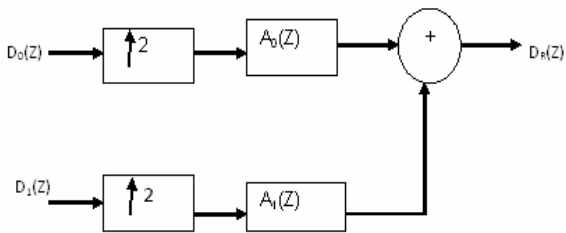

Fig. 5 Synthesis bank of wavelet structure

Fig.5 shows synthesis bank of wavelet structure. $d_0(n)$ is low pass input signal, $d_1(n)$ is high pass input signal and $d_R(n)$ is reconstructed signal

For simplicity write in Z domain

$$D_R(Z) = A_0(Z)D_0(Z^2) + A_1(Z)D_1(Z^2) \tag{13}$$

From Quadrature Mirror Filter by [8] at perfect reconstruction, filters are chosen as follows

$$A_0(Z) = 2B(Z) \leftrightarrow 2b(n) \tag{14}$$

$$A_1(Z) = -A(-Z) = -2B(-Z) \leftrightarrow 2(-1)^{n+1}b(n) \tag{15}$$

Refer to Eq (7)

A(Z) is decomposed into

$$A(Z) = \sum_{\lambda=0}^{M-1} Z^{-\lambda} A\lambda(Z^M) \tag{16}$$

In Haar Wavelet M=2

$$A_0(Z) = A_{00}(Z^2) + z^{-1}A_{01}(Z^2) \tag{17}$$
$$A_1(Z) = -A_{00}(Z^2) + z^{-1}A_{01}(Z^2) \tag{18}$$

Sub Eq. 17 & 18 in (13)

$$D_R(Z) = D_0(Z^2)[A_{00}(Z^2) + z^{-1}A_{01}(Z^2)] + [-A_{00}(Z^2) + z^{-1}A_{01}(Z^2)]D_1(Z^2)$$

$$D_R(Z) = A_{00}(Z)[D_0(Z^2) - D_1(Z^2)] + z^{-1}A_{01}(Z^2)[D_0(Z^2) + D_1(Z^2)] \tag{19}$$

Up sampler at the input of the synthesis filter bank will moved to output. So Eq.(19) can be drawn by

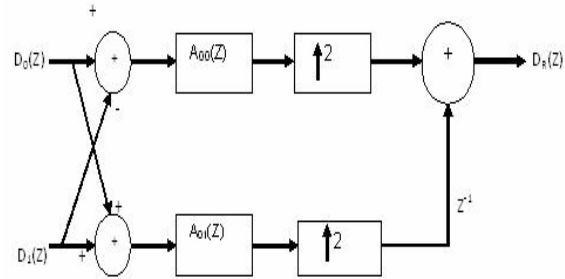

Fig. 6 Modified synthesis bank structure

In Haar wavelet $A_{00}(Z) = A_{01}(Z) = B_{00}(Z)$

In Haar wavelet $b_{00}(n) = a_{00}(n) = \frac{1}{\sqrt{2}}$.

Draw in time domain

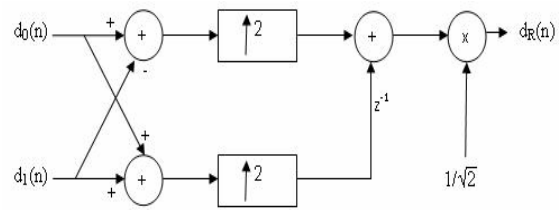

Fig. 7 Fast Haar wavelet synthesis bank

Combining Fig.4 & Fig.7, Fast Haar Wavelet Structure is obtained. Compared to Fig.2, Number of Mathematical calculations are reduced in Fast Haar Wavelet Structure is shown in Fig.8.

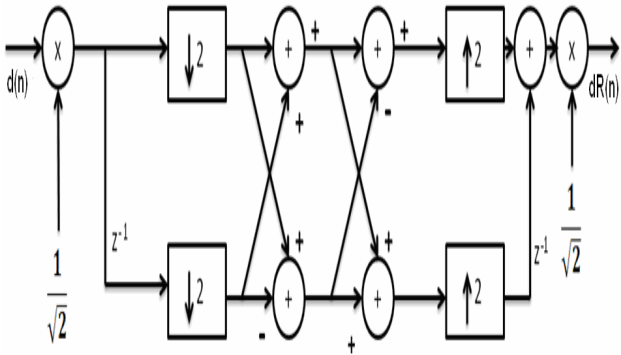

Fig. 8 Fast Haar wavelet structure

## V. EXPERIMENTAL RESULTS

The results of applying, for one subject, which the signal is taken from laser based noninvasive Doppler indigenous developed equipment, the novel Fast Haar wavelet with approximation data are shown in Fig.9 shows that difference between original haar wavelet and Fast haar wavelet are matched well. The Error rate between existing and proposed Fast Haar wavelet at -90dB are shown in Fig. 11.

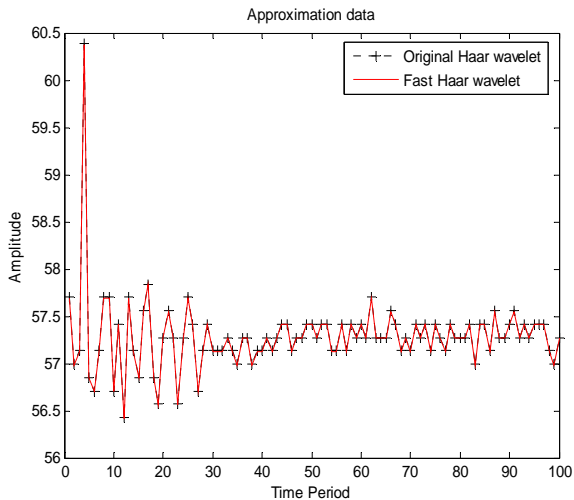

Fig. 9 Results of approximation data compared to existing and Proposed Fast Haar wavelet Transform.

Similarly from the same novel Fast Haar wavelet with detail data are shown in Fig.10 shows that difference between original Haar wavelet and Fast Haar wavelet are matched well. The Error rate between existing and proposed Fast Haar wavelet at -160dB to -220dB are shown in Fig.11.

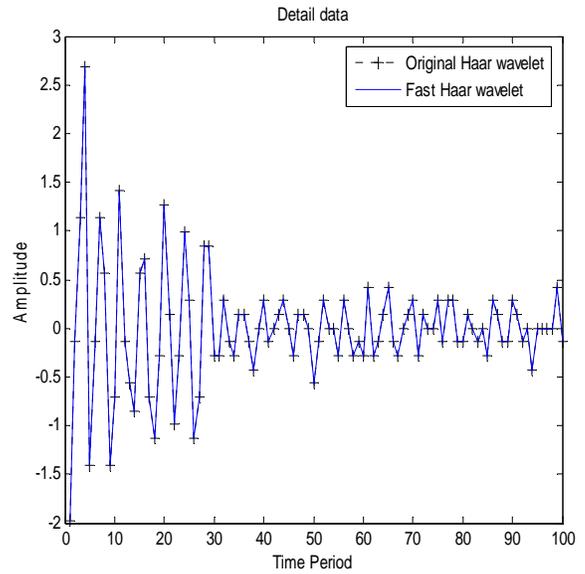

Fig. 10 Results of detail data compared to existing and Proposed Fast Haar wavelet Transform.

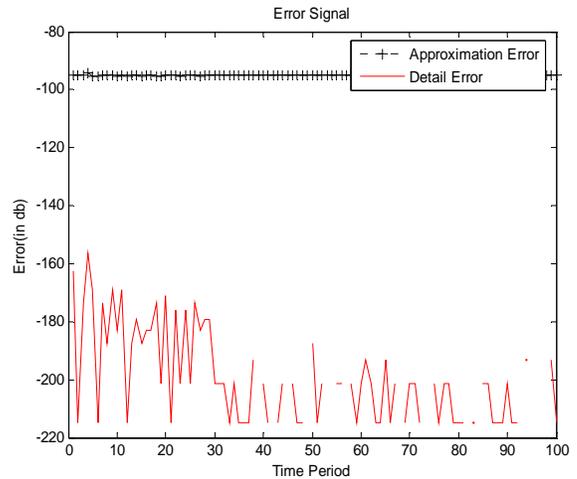

Fig. 11 Results of Error rate compared to existing and Proposed Fast Haar wavelet Transform

We have checked our proposed method in image processing also. Lowpass output was obtained by applying original Haar wavelet and proposed Fast Haar wavelet. Fig.12(a) shows Lena image, Fig.12(b) shows lowpass image of lena by applying original Haar wavelet transform and Fig.12(c) shows lowpass image by applying Fast Haar wavelet transform. Fig.12(d) shows difference between Fig.12(b) & Fig.12(c) From the Fig.12(d), it is clearly visible difference value for all coefficients are less.

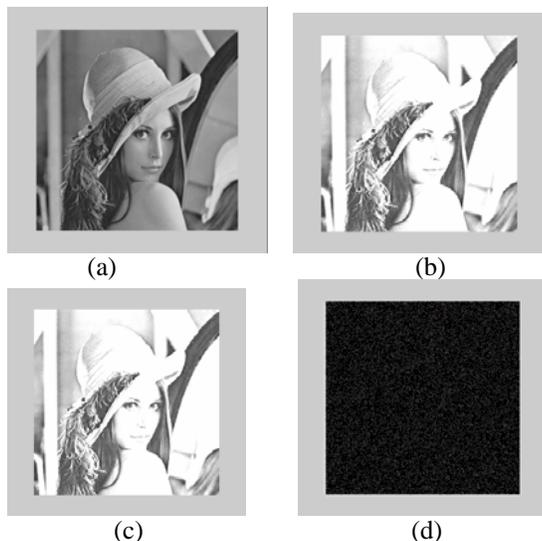

(a)     (b)

(c)     (d)

Fig.12 Comparison of Fast haar wavelet with original Haar wavelet
a) Lena image (b) Lowpass of Lena image by original Haar wavelet
(c) Lowpass of Lena image by Fast Haar wavelet
(d) Difference between lowpass output by original Haar wavelet & Fast Haar wavelet

## VI. CONCLUSION

This work presents a novel Fast Haar wavelet estimator, for application to biosignals such as noninvasive doppler signals and medical images. . In this paper, signals and images are decomposed and reconstructed by Haar wavelet transform without convoution. The proposed method allows for the dynamic reduction of power and computational complexity than the conventional method. The error rate between the conventional and the proposed method was reduced in the signal and image procesing.

## AUTHORS PROFILE

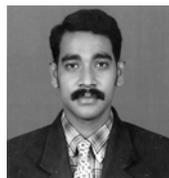

Mr.V.Ashok received the Bachelors degree in Electronics And Communication Engineering from Bharathiyar University, Coimbatore in 2002 and the Master degree in Process Control And Instrumentation Engineering form Annamalai University, Chidambaram in 2005. Since then, he is working as a Lecturer in Velalar College of Engineering and Technology (Tamilnadu), India. Presently he is a Part time (external) Research Scholar in the Department of Electrical Engineering at Anna University, Chennai (India). His fields of interests include Medical Electronics, Process control and Instrumentation and Neural Networks.

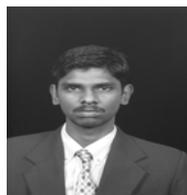

Mr.T.Balakumaran received the Bachelors degree in Electronics and Communication Engineering from Bharathiyar University, Coimbatore in 2003 and the Master degree in Applied Electronics from Anna University, Chennai in 2005. Since then, he is working as a Lecturer in Velalar College of Engineering and Technology (Tamilnadu), India. Presently he is a Part time (external) Research Scholar in the Department of Electrical Engineering at Anna University, Coimbatore (India). His fields of interests include Image Processing, Medical Electronics and Neural Networks.

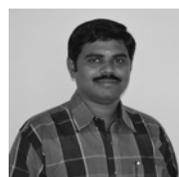

Mr.C.Gowri Shankar received the B.E Electrical and Electronics Engineering from Periyar University in 2003 and M.E Applied electronics from Anna University, Chennai in 2005. Since 2006, he has been a Ph.D. candidate in the same university. His research interests are Multirate Signal Processing, Computer Vision, Medical Image Processing, and Pattern Recognition. Currently, he is working in Dept of Electrical and Electronics Engineering, Velalar College of Engineering and Technology, Erode.

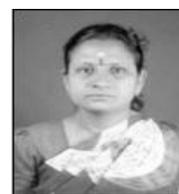

Dr.ILA.Vennila received the B.E Degree in Electronics and Communication Engineering from Madras University, Chennai in 1985 and ME Degree in Communication System from Anna university, Chennai in 1989. She obtained Ph. D. Degree in Digital Signal Processing from PSG Tech, Coimbatore in 2006. Currently she is working as Assistant Professor in EEE Department, PSG Tech and her experience started from 1989; she published about 35 Research Articles in National, International Conferences National and International journals. Her area of interests includes Digital Signal Processing, Medical Image processing, Genetic Algorithm and fuzzy logic

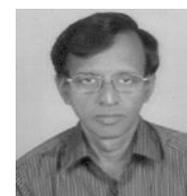

Dr.A.Nirmalkumar. A, received the B.Sc.(Engg.) degree from NSS College of Engineering, Palakkad in 1972, M.Sc.(Engg.) degree from Kerala University in 1975 and completed his Ph.D. degree from PSG Tech in 1992. Currently, he is working as a Professor and Head of the Department of Electrical and Electronics Engineering in Bannari Amman Insititute of Technology, Sathyamangalam, Tamilnadu, India. His fields of Interest are Power quality, Power drives and control and System optimization.